\documentclass[prc,preprint,tightenlines,nofootinbib]{revtex4}

\usepackage{graphicx}  
\usepackage{bm}  
\usepackage{amsmath}
\usepackage{epsf}
\newcommand{\beq}{\begin{equation}}
\newcommand{\eeq}{\end{equation}}
\newcommand{\bea}{\begin{eqnarray}}
\newcommand{\eea}{\end{eqnarray}}

\newcommand{\lsim}{\, \, \raisebox{-0.8ex}{$\stackrel{\textstyle <}{\sim}$ }}
\newcommand{\simlt}{\stackrel{<}{{}_\sim}}

\begin{document}
\title{The Three-Boson System at Next-To-Next-To-Leading Order}
\author{L. Platter}\email{lplatter@phy.ohiou.edu}
\author{D.~R. Phillips}\email{phillips@phy.ohiou.edu}
\affiliation{Department of Physics and Astronomy, Ohio University,
Athens, OH 45701, USA}
\date{\today}
\begin{abstract}
We discuss effective field theory treatments of the problem of three
particles interacting via short-range forces (range $R \ll a_2$, with
$a_2$ the two-body scattering length).  We show that forming a
once-subtracted scattering equation yields a scattering amplitude
whose low-momentum part is renormalization-group invariant up to
corrections of $O(R^3/a_2^3)$. Since corrections of $O(R/a_2)$ and
$O(R^2/a_2^2)$ can be straightforwardly included in the integral
equation's kernel, a unique solution for 1+2 scattering phase shifts
and three-body bound-state energies can be obtained up to this
accuracy. We use our equation to calculate the correlation between the
binding energies of Helium-4 trimers and the atom-dimer scattering
length.  Our results are in excellent agreement with the recent
three-dimensional Faddeev calculations of Roudnev and collaborators
that used phenomenological inter-atomic potentials.
\end{abstract}
\maketitle
\section{Introduction}
\label{sec:intro}

Zero-energy scattering between two particles can be characterized by
one number: the scattering length $a_2$.  The zero-energy
cross-section for scattering of two non-relativistic
quantum-mechanical particles is:
\begin{equation}
\sigma=4 \pi a_2^2.
\label{eq:zeroenxsn}
\end{equation}
This formula governs the cross section at center-of-mass momenta $k$
which are significantly smaller than $1/R$, where $R$ is the range of
the underlying two-particle potential. In the domain where $k R \ll 1$
an effective field theory (EFT) with contact interactions alone can be
used to organize the corrections to Eq.~(\ref{eq:zeroenxsn}), and 
relate $\sigma$ to other observables, via a controlled expansion in a small
parameter or small parameters.  The nature of this expansion is,
however, quite different depending on the relative sizes of the
scattering length $a_2$ and the range $R$.  In the first case, $a_2
\sim R$, one can calculate corrections to Eq.~(\ref{eq:zeroenxsn}) via
a perturbative expansion in $a_2\,k$. But, if $a_2 \gg R$---as is true
in nuclear physics and a number of other quantum-mechanical
systems---the expansion in $k a_2$ breaks down very quickly. It is then
more efficient to develop an expansion in the small parameters
$R/a_2$ and $k \, R$.  Key to this modified expansion is that, if
$a_2 > 0$, the large scattering length is associated with a shallow
bound state with binding energy 
\beq 
B_2=\frac{1}{m a_2^2}~.  
\label{eq:B2LO}
\eeq
(Here and in the rest of this work we set $\hbar=1$.) The existence of
this bound state signals the onset of non-perturbative physics in the
EFT. In order to describe physics at $k \sim 1/a_2 \ll 1/R$ we need an
EFT expansion that already at its leading order
generates the two-body bound state with energy
$B_2$ (or its analog on the second Riemann sheet if $a_2 < 0$). The
properties of this bound state can then be accessed by scattering
particles from the bound state, i.e. examining the three-body
scattering problem which we will refer to from now on as ``1+2
scattering''. It is this type of scattering which we will focus on in
this paper.

The most extreme example of the situation $a_2 \gg R$ occurs when we
take the limit $a_2 \rightarrow \infty$, or, equivalently $R
\rightarrow 0$. When zero-range interactions are involved the
three-body problem is ill-defined without the introduction of an
additional length scale, which can be chosen to be the scattering
length for the 1+2 process,
$a_3$~\cite{STM,Danilov,Bedaque:1998kg}. This means that in the limit
$a_2 \rightarrow \infty$ the three-body system is characterized solely
by the ratio $a_3/a_2$. Universal predictions for the low-energy
(i.e. $k R \ll 1$) behavior of any system with a particular value of
that ratio can then be generated---either using an EFT or
quantum-mechanical models. Since the EFT generates predictions for
these systems which do not depend on the details of the short-distance
physics it provides a powerful way of unifying apparently diverse
physics problems: dimers, trimers 
and tetramers of Helium-4 atoms; deuterium, tritium, \& Helium nuclei; and nuclei
with neutron halos~\cite{Bedaque:2002yg,Platter:2004qn,Platter2005,Bertulani:2002sz,Yamashita:2004pv}. 
These systems
differ in size and spin structure but share the same essential feature
of a large two-body scattering length.  For a recent review of this
physics, see Ref.~\cite{Braaten:2004rn}.

In reality, no system can exist exactly in the universality limit $a_2
\rightarrow \infty$. Being able to calculate the deviations from, and
approach to, universality is important for all such problems. That
approach is controlled by the parameter $R/a_2$. A first assessment of
the impact that taking $R/a_2 \neq 0$ has on universality predictions
was carried out by Efimov \cite{Ef00,Ef01}, and these calculations
were recently systematized using EFT by Hammer \&
Mehen~\cite{Hammer:2001gh} and Bedaque {\it et
  al.}~\cite{Bedaque:2002yg}, both of whom showed that the $O(R/a_2)$
corrections to universality came from the physics of two-body
scattering alone. This means that, up to corrections of
$O(R^2/a_2^2)$, the low-energy properties of the three-body system are
determined by two numbers: the ratios $R/a_2$ and $a_3/a_2$.

In this paper we take this argument one step further, and use EFT to
show that the properties of the three-body system are actually
determined by those two ratios {\it up to corrections of
  $O(R^3/a_2^3)$}. In order to do this we develop the integral
equation describing three-body scattering up to an arbitrary order in
$R/a_2$. It has been known for forty years that to zeroth order in
this parameter, i.e. when $a_2 \rightarrow \infty$, this integral
equation is ill-defined~\cite{Danilov}. In Section~\ref{sec:sublo} we
discuss the nature of this problem in the language of the
renormalization group, and reiterate the argument of
Ref.~\cite{Afnan:2003bs} in order to show that the equation can be
made well-defined by one subtraction, which depends on the three-body
parameter $a_3$.  Therefore once $a_2$ and $a_3$ are known the
integral equation may be used to generate the predictions for 1+2
scattering phase shifts, three-body bound-state energies,
recombination rates, etc. mentioned above. Then, in
Section~\ref{sec:3boson} we show that this same subtraction also makes
the equation well-defined when $O(R/a_2)$ and $O(R^2/a_2^2)$
corrections are included in the kernel of the integral equation. This
suffices to prove that only $a_2$, $a_3$, and $R$ are needed to
generate predictions for three-body system observables which are
accurate up to $O(R^3/a_2^3)$. Bedaque {\it et
  al.}~\cite{Bedaque:2002yg} and Grie\ss
hammer~\cite{Griesshammer:2005ga} have claimed that an additional
three-body datum is necessary to achieve this accuracy, but our
results show that renormalization does not necessitate the inclusion
of this information in the calculation at $O(R^2/a_2^2)$.

These results are true for all of the physical systems listed
above---indeed, for any system for which $R/a_2$ is a small
parameter. In order to demonstrate their utility we focus on an
apparently straightforward application: to dimers and trimers of
Helium-4 atoms.  In Section~\ref{sec:results} we use our formalism to
generate predictions for the binding energies of trimers of Helium-4
atoms. Since the atom-atom scattering length is
$a_2=\left(104^{+8}_{-18}\right)$\AA\, while the typical van der
Waal's forces between the Helium atoms have a range $R\approx 7$~\AA,
in this case the expansion parameter $R/a_2 \simlt 10$\%.  The $^4$He
trimer, tetramer, and several larger $^4$He clusters have been
observed \cite{STo96,BST02}, but unfortunately there are no
measurements of trimer binding energies, and so we use precise
few-body calculations~\cite{NFJ98,RoY00,Ro00,MSSK01,Barl02}, based on
quite complicated inter-atomic potentials~\cite{TTYref,HFDBref}, to
provide the two input ratios for our EFT. We then compare the EFT's
predictions with the corresponding results obtained in
Refs.~\cite{MSSK01,Ro00}. In fact, the numbers obtained for the
atom-dimer scattering length in these two papers differ by
more than the stated systematic error of the computations. Our EFT
calculation confirms the result of Ref.~\cite{Ro00}, but we are unable
to reproduce the result for the scattering length obtained in
Ref.~\cite{MSSK01}. Indeed, the discrepancy with their number is two
orders of magnitude larger than the expected $R^3/a_2^3$ error of our
calculation. From the perspective of the EFT this renders the results
of Ref.~\cite{MSSK01} suspect, since they are not in accord with the
model-independent pattern of $R/a_2$ corrections that we have computed
here.

With the addition of spin and isospin degrees of freedom, the
formalism developed here can be applied to neutron-deuteron and (with
the inclusion of Coulomb corrections) proton-deuteron scattering at
low energies. In fact, the power counting used here for atomic systems
has been quite thoroughly tested in the nuclear context. Calculations
to NNLO (and several orders beyond in the $NN$ system) converge as
expected~\cite{Rupak,CBK,Bedaque:2002yg}. We will pursue the use of
our formalism for such systems, and also develop a perturbative
expansion of the solution of our integral equation, in future work.

\section{Three-Body Scattering at Leading Order}
\label{sec:sublo}
In this section we will give a short introduction to the application
of the EFT with contact interactions alone to the three-body sector
and the renormalization of the integral equation that results after one subtraction.

At very low energies all short-range interactions appear point-like and
one can describe the physics of a system that contains only such
interactions using an EFT built up from contact terms alone. The most
general Lagrangian describing a non-relativistic system of identical
bosons with short-range interactions is given by \bea
\label{eq:lagrangian1}
\mathcal{L}& =&{\psi}^{\dagger}\left(i
  \partial_t+\frac{\overrightarrow{\nabla}^2}{2m}\right)\psi
-\frac{C_0}{2}(\psi^\dagger\psi)^2
-\frac{D_0}{6}(\psi^{\dagger}\psi)^3+ \ldots,
\eea
where the ellipses denote interactions with more derivatives
and/or more fields.
For the application to the three-body sector it is useful to rewrite this
Lagrangian in terms of a particle field and a {\it dimeron} field, $T$, which has
the quantum numbers of a two-body bound state
\bea
\mathcal{L}&=&{\psi}^{\dagger}\left(i
  \partial_t+\frac{\overrightarrow{\nabla}^2}{2m}\right)\psi
+\Delta T^\dagger T-\frac{g}{\sqrt{2}}(T^\dagger\psi\psi+\hbox{h.c.})
+h T^\dagger T\psi^\dagger\psi \ldots .
\label{eq:lagrangian2}
\eea
It can be demonstrated~\cite{Ka97,BG98} that these terms
give the leading-order (LO) t-matrix for two-to-two scattering, in the
context of the EFT expansion in powers of $R/a_2$ developed in 
Refs.~\cite{vK97,Ka98,Ge98,Ri98}. In the
normalization we use in this paper that LO t-matrix is:
\begin{equation}
\tau^{(0)}(E)=\frac{2}{\pi m^2} \frac{\gamma + \sqrt{- m E}}{E + B_2} \equiv
\frac{S^{(0)}(E)}{E+B_2},
\label{eq:LOtau}
\end{equation}
where we have adjusted the constants $g$ and $\Delta$ such that
the two-body binding energy $B_2$ is correctly reproduced, and:
\beq
S^{(0)}(E)=\frac{2}{\pi  m^2}\left[\gamma+\sqrt{-mE}\right]~.
\eeq
Here, and in what follows, the superscript ${}^{(0)}$ indicates that
this is a leading order result. Also $\gamma=\sqrt{m B_2}$ is the
``binding momentum'' of the two-body system. Equation~(\ref{eq:B2LO})
implies that at LO  $\gamma=\frac{1}{a_2}$.

When embedded in the three-body system the two-body amplitude
(\ref{eq:LOtau}) generates the following equation for the $1+2$
scattering amplitude, $X_\ell$, provided that the three-body force $h$
is, for the time being, ignored:
\beq
\label{eq:Xl}
X_{\ell}^{(0)}(q,q';E)=2 Z_{\ell}(q,q';E)+\int_0^\Lambda\hbox{d}q'' q''^2
\,2Z_{\ell}(q,q'',E)\frac{S^{(0)}(E;q'')}
{E-\frac{3}{4}\frac{q''^2}{m}+B_2}X_{\ell}(q'',q';E)~,
\eeq
where $\ell$ denotes the relative angular momentum between the dimer
and the third particle and $S^{(0)}(E;q'') \equiv S^{(0)}(E-\frac{3q''^2}{4m})$ .

Considering only $s$-waves and defining
\beq
2 Z_0(q,q';E)=\mathcal{Z}(q,q';E)=-\frac{m}{q q'}
\log\left(\frac{q^2+q'^2+q q'-mE}{q^2+q'^2-q q'-mE}\right)~,
\eeq
we reformulate the integral equation in Eq.~(\ref{eq:Xl})
via a principal-value integral
\beq
\label{eq:kmatrix}
K^{(0)}(q,q';E)=\mathcal{Z}(q,q';E)+\mathcal{P}\int_0^\Lambda\hbox{d}q''\,q''^2
\mathcal{Z}(q,q'',E)\frac{S^{(0)}(E;q'')}{E+B_2-\frac{3}{4}q''^2}
K^{(0)}(q'',q';E)~.
\eeq
Below breakup it is convenient to use this $K$-matrix approach.
(From now on we drop the subscript $\ell$, in order to avoid
notational  clutter. For the rest of the paper we will focus only on 
$\ell=0$.)
The $K$-matrix is normalized such that it gives on-shell ($E=\frac{3
  k^2}{4 m} - B$):
\beq
K^{(0)}(k,k;E)=-\frac{3 m}{8\gamma k}\tan\delta~.
\eeq
Using the boson-dimer effective-range expansion for the s-wave phase shift
\beq
k\cot\delta=-\frac{1}{a_3}+\frac{1}{2}k^2 r_3+...~,
\eeq
with $a_3$ and $r_3$ being respectively, the 1+2
scattering length and effective range, we find
\beq
K^{(0)}(0,0;-B_2)=\frac{3 m a_3}{8\gamma}~.
\label{eq:Kthreshon}
\eeq
The solution to the integral equation (\ref{eq:Xl}) is strongly cutoff
dependent.  This is a manifestation of the fact that Eq.~(\ref{eq:Xl})
is not properly renormalized. Bedaque, Hammer, and van
Kolck~\cite{Bedaque:1998kg} have shown that one way to remedy this
problem is to insert the dimer-boson coupling $h$ of
Eq.~(\ref{eq:lagrangian2}) into the kernel of the leading-order
equation. This makes it possible to fix the behavior of the amplitude
for arbitrary cutoffs by adjusting $h$ such that a chosen three-body
observable is reproduced correctly.

An equivalent procedure is to use subtracted equations. In this approach
we trade the three-body force $h$ for a three-body observable which
appears explicitly in the equation. The result is a formulation of the
three-body problem with short-range forces in which only renormalized
quantities appear.

We begin by considering:
\beq
K^{(0)}(q,0;-B_2)=\mathcal{Z}(q,0;-B_2)-\frac{4m}{3}\int_0^\Lambda\hbox{d}
q''\,\mathcal{Z}(q,q'';-B_2)S^{(0)}(-B_2,;q'')K^{(0)}(q'',0;-B_2)~,
\label{eq:threshoffshell}
\eeq
We can insert the required three-body input by demanding
that the scattering amplitude reproduces the correct particle-dimer
scattering length at threshold. The on-shell amplitude at $E=-B_2$
is given by Eq.~(\ref{eq:Kthreshon}) and it obeys
\bea
\label{eq:threshonshell}
K^{(0)}(0,0;-B_2)=\mathcal{Z}(0,0;-B_2)
\nonumber -\frac{4m}{3}\int_0^\Lambda\hbox{d}
q''\,\mathcal{Z}(0,q'';-B_2)S^{(0)}(-B_2,;q'')K^{(0)}(q'',0;-B_2)~,
\nonumber\\
\eea
where $a_3$ denotes the atom-dimer scattering length. By subtracting
Eq.~(\ref{eq:threshonshell}) from Eq.~(\ref{eq:threshoffshell}) and
using Eq.~(\ref{eq:Kthreshon}) we obtain
the subtracted equation at threshold\footnote{Note that we would still
 have obtained Eq.~(\ref{eq:threshsubt}) if we had started out with a scattering
equation containing the momentum-independent three-body force $h$ and 
considered the equations corresponding to (\ref{eq:threshoffshell})
and (\ref{eq:threshonshell})  in that case.}:
\bea
K^{(0)}(q,0;-B_2)&=&\frac{3ma_3}{8 \gamma}
+\Delta[\mathcal{Z}](q,0;-B_2)
\nonumber\\[0.3cm]
&&\quad-\frac{4m}{3}\int_0^\Lambda\hbox{d}q''
\Delta[\mathcal{Z}](q,q'';-B_2)
S^{(0)}(-B_2;q'')K^{(0)}(q'',0;-B_2)~,
\label{eq:threshsubt}
\eea where \beq
\Delta[\mathcal{Z}](q,q';E)=\mathcal{Z}(q,q';E)-\mathcal{Z}(0,q';E)~.
\eeq This equation was first derived in a different notation by Hammer
and Mehen~\cite{Hammer:2000nf}. As they showed, it is
renormalization-group invariant up to corrections which scale $\sim
q^2$---a result that is in accord with the more recent analysis of
Ref.~\cite{BB05}.

Now we can determine the full-off-shell amplitude at threshold by exploiting
the symmetries of $K$. Before subtraction
the full off-shell amplitude at threshold satisfies
\beq
\label{eq:full-off-shell-not-renormalized}
K^{(0)}(q,q';-B_2)=\mathcal{Z}(q,q';-B_2)
-\frac{4m}{3}\int_0^\Lambda\hbox{d}q''\,\mathcal{Z}(q,q'';-B_2)
S^{(0)}(-B_2;q'')K^{(0)}(q'',q';-B_2)
\eeq
Since
$\mathcal{Z}(q,q';E)=\mathcal{Z}(q',q;E)$, we also have
$K^{(0)}(0,q;-B_2)=K^{(0)}(q,0;-B_2)$. 
Using this fact as input, together with the solution of
Eq.~(\ref{eq:threshsubt}), allows the subtracted version of
Eq.(\ref{eq:full-off-shell-not-renormalized}):
\bea
\nonumber
K^{(0)}(q,q';-B_2)&=&K^{(0)}(0,q';-B_2)+\Delta[\mathcal{Z}](q,q';-B_2)\\[0.3cm]
&&\quad-
\frac{4 m}{3}\int_0^\Lambda\hbox{d}q'' 
\Delta [\mathcal{Z}](q,q'';-B_2)S^{(0)}(-B_2;q'')K^{(0)}(q'',q';-B_2)~,
\label{eq:fulloffshell}
\eea
to determine the fully-off-shell amplitude at threshold.

With Eq.~(\ref{eq:fulloffshell}) in hand resolvent identities may be
used to obtain the subtracted amplitude at any energy~\cite{Afnan:2003bs}:
\beq
\label{eq:full-amplitude-subtracted}
K^{(0)}(q,k;E)=K^{(0)}(q,k;-B_2)+B^{(0)}(q,k;-B_2)+
\int_0^\Lambda\hbox{d}q' q'^2\,Y^{(0)}(q,q';E) K^{(0)}(q',k;E)~,
\eeq
where the second inhomogeneous term is given by
\beq
B^{(0)}(q,k;E)=\delta[\mathcal{Z}](q,k;E)+\int_0^\Lambda\hbox{d}q''\,{q''}^2\,
K^{(0)}(q,q'';-B_2) \, \tau^{(0)}(-B_2;q'')\delta[\mathcal{Z}](q'',k;E)~,
\eeq
with
\beq
\delta[\mathcal{Z}](q,q';E)=\mathcal{Z}(q,q';E)-\mathcal{Z}(q,q';-B_2)~,
\eeq
and $\tau(E;q)=\tau(E-{\textstyle\frac{3q^2}{4m}})$. The kernel in
Eq.(\ref{eq:full-amplitude-subtracted}) is
\bea
\nonumber
Y^{(0)}(q,q';E)&=&K^{(0)}(q,q';-B_2)\delta[\tau^{(0)}](E;q')
+\delta[\mathcal{Z}](q,q';E)
\tau^{(0)}(E;q')\\
\nonumber
&&\qquad+\int_0^\Lambda\hbox{d}q'' \,q''^2\,K^{(0)}(q,q'';-B_2)
\tau^{(0)}(-B_2;q'')
\delta[\mathcal{Z}](q'',q';E)\tau^{(0)}(E;q')\\
&=&K^{(0)}(q,q';-B_2)\delta[\tau^{(0)}](E;q')+B^{(0)}(q,q';E)\tau^{(0)}(E;q')~,
\label{eq:Y0}
\eea
with:
\beq
\delta[\tau^{(0)}](E;q')=\tau^{(0)}(E;q') - \tau^{(0)}(0;q').
\label{eq:deltatau}
\eeq

The asymptotic behavior of the elements of
Eqs.~(\ref{eq:full-amplitude-subtracted})--(\ref{eq:deltatau}) is as follows:
\begin{eqnarray}
\tau^{(0)}(E;q') \sim \frac{1}{q'}~&\mbox{for}&~q' \gg k \label{eq:taubeh}\\
\delta[\tau^{(0)}](E;q') \sim \frac{k^2}{q'^3}~&\mbox{for}&~q'
\gg k\\
K^{(0)}(q,q';-B_2) \sim \frac{1}{qq'}~&\mbox{for}&~q, q' \gg \gamma\\
K^{(0)}(q,q';-B_2) \sim \frac{1}{q \gamma}~&\mbox{for}&~q' \sim
\gamma, 
\quad q \gg
q', \gamma\\
\delta[{\cal Z}](q,q';E) \sim \frac{k^2}{q^2q'^2}~&\mbox{for}&~q \sim q' \gg
k, \gamma\label{eq:deltaZ}
\end{eqnarray}
(In Eqs.~(\ref{eq:taubeh})--(\ref{eq:deltaZ}) we omit factors of
$m$ since they are simply overall factors in the final answers obtained for these 
non-relativistic systems.)
In the case of $K^{(0)}$ the behavior listed is
modulated by an oscillatory function, but that is not important for
our purposes here. What is important is that
Eqs.~(\ref{eq:taubeh})--(\ref{eq:deltaZ}) result in
\begin{equation}
K^{(0)}(q,k;E) \sim \frac{1}{q \gamma}~\mbox{for}~q \gg \gamma,k;
\quad k \sim \gamma.
\end{equation}

This means that the
Eqs.~(\ref{eq:full-amplitude-subtracted})--(\ref{eq:deltatau}) are
well behaved, in the sense that for $\Lambda \gg k, \gamma$ the
results are independent of $\Lambda$---up to corrections of relative
order $\frac{q^3}{\Lambda^3}$. This is equivalent to the statement
that Eq.~(\ref{eq:Xl}) can be renormalized---i.e. made to render
cutoff-independent predictions---by the introduction of a
momentum-independent force like the one written explicitly in
Eq.~(\ref{eq:lagrangian1}). But the set of equations developed here
(following Refs.~\cite{Afnan:2003bs,Hammer:2000nf}) allows us to
compute three-body phase shifts without the inclusion of an explicit
three-body force. 

Furthermore, the energies at which the homogeneous
version of Eq.~(\ref{eq:full-amplitude-subtracted}):
\begin{equation}
K^{(0)}(q,k;-B_3)=
\int_0^\Lambda\hbox{d}q' \, q'^2\,\,Y^{(0)}(q,q';-B_3) K^{(0)}(q',k;-B_3)~,
\label{eq:homog}
\end{equation}
has a non-trivial
solution will be the energies at which three-body bound states
occur. In fact, there are infinitely many such energies, most of which
lie outside the domain of validity of the EFT, and so should be
regarded as spurious solutions of Eq.~(\ref{eq:homog}). However, those
eigenenergies $B_3$ which are within the radius of convergence of the
theory represent predictions which should be reliable within the
error estimate at the order under consideration (leading order in this
case). The EFT equations developed here therefore explicitly show that
there will be correlations between the three-body scattering length
$a_3$ and bound-state energies which are $\lsim 1/mR^2$.
\section{The Subtracted Equations at NNLO}
\label{sec:3boson}

In this section we will discuss how to include corrections which are
higher order in the $R/a_2$ expansion using the subtraction formalism. We
will show that up to next-to-next-to-leading order in this expansion
(NNLO) no additional three-body counterterm is needed for consistent
renormalization.

Whether a new counterterm is needed or not at a particular order of an
EFT can usually be determined with the help of naive dimensional
analysis (NDA). In perturbative EFTs such as chiral perturbation
theory this sort of analysis allows us to predict at what order terms of a
given structure must be included in the Lagrangian in order to
renormalize the theory.

However, in EFTs where non-perturbative resummation of a particular
set of diagrams is mandated by the power counting, NDA may not be a
reliable guide to the size of higher-dimensional operators. For
instance, in the EFT that we are using here, and whose Lagrangian is
given by Eq.~(\ref{eq:lagrangian1}), NDA leads us to expect that the
size of the three-body force $D_0$ will be $R^5/M$. This would make
that three-body force an effect that only needed to be considered at
NNLO.  However, Bedaque {\it et al.} found it necessary to include
such a three-body force at leading order, so as to render three-body
observables cutoff independent~\cite{Bedaque:1998kg}. In three-body
systems with $a_2 \gg R$ renormalization forces the three-nucleon
coupling $D_0$ to have a larger size than we would expect based on NDA
applied at the momentum scale $1/R$.

The authors of Ref.~\cite{Bedaque:2002yg} demanded that with each
further order included in the calculation the corrections should scale
with an additional power of $R/a_2$. Using analytical and numerical
arguments they concluded that an additional three-body force enters at
NNLO. A renormalization-group analysis of the perturbations around the
limit cycle found in Ref.~\cite{Bedaque:1998kg} appears to support
this view~\cite{BB05}. In contradistinction to the approach of
Ref.~\cite{Bedaque:2002yg}, here we let NDA and the question of
renormalizability be the ultimate guides to power counting in the
three-body problem with short-range interactions. A new structure
should be included in the effective Lagrangian of the theory if it is
necessary for renormalization, i.e. it must be included so as to make
observables cutoff independent up to the order under consideration,
{\it or} if its presence is implied at that order by NDA with respect
to the heavy scales in the theory. These criteria have the advantage
that they are not subjective. They therefore allow the EFT
(\ref{eq:lagrangian1}) to make testable predictions about data in
systems in which there are only short-range forces.  If, while using
the number of short-distance parameters that is indicated at a given
order by our combined criteria of renormalizability and NDA, the EFT
(\ref{eq:lagrangian1}) is unable to describe data with the accuracy
expected at that order then we can only conclude that the forces at
work in that system are not truly short-ranged.

Analysis in this spirit beyond leading order is most transparent if a
perturbative expansion of the scattering amplitude (or, equivalently,
the K-matrix) in powers of $R/a_2$ can be established. In
Ref.~\cite{Hammer:2001gh} Mehen and Hammer showed that in such an
expansion no three-body force beyond the $h$ term in the leading-order
computation of Ref.~\cite{Bedaque:1998kg} was needed to renormalize
the $O(R/a_2)$ piece of the scattering amplitude.

However, it should also be possible to imitate the approach of
Weinberg in the nuclear effective theory with
pions~\cite{Weinberg:1990,Weinberg:1991} and expand the kernel of the
integral equation governing 1+2 scattering in powers of $R/a_2$. If we
proceed in this way we are certainly not doing a calculation that is
less accurate than the perturbative one discussed in the
previous paragraph---provided we can properly renormalize the
resulting amplitudes. In fact, here we will show that if we calculate
the kernel including terms of order $R/a_2$ and $R^2/a_2^2$ then the
renormalization already displayed for the leading-order calculation in
Section~\ref{sec:sublo} is sufficient to facilitate predictions for 1+2
scattering. In other words, the same subtraction suffices to
renormalize the theory at leading, next-to-leading, and
next-to-next-to-leading order.

To develop the integral equation that governs 1+2 scattering
at higher orders in the $R/a_2$ expansion we first note that 
the full two-body t-matrix with effective-range contributions 
resummed is given by~\cite{Bethe,BS00}
\beq
\tau(E)=-\frac{2}{\pi m}\,\frac{1}{-\gamma+\sqrt{-mE}
+\frac{r}{2}\left(\gamma^2+mE\right)}~.
\label{eq:TBamp}
\eeq 
Corrections to Eq.~(\ref{eq:TBamp}) involve the shape parameter,
and, for natural values of that parameter, are suppressed by
$(R/a_2)^3$.  The $\tau$ of Eq.~(\ref{eq:TBamp}) has poles
which represent bound states with energies outside the region of
validity of the EFT. It therefore
cannot be used in the integral equation for cutoffs $\Lambda >
1/r$ unless further effort is devoted to the subtraction of these
unphysical bound states.  Therefore, we expand the dimeron propagator
(\ref{eq:TBamp}) up to a given order in $R/a_2$. The leading-order
result was already displayed in Eq.~(\ref{eq:LOtau}). At higher orders
we obtain \beq \tau^{(n)}(E)=\frac{S^{(n)}(E)}{E+B_2}, \eeq with: \beq
S^{(n)}(E)=\frac{2}{\pi m^2} \sum_{i=0}^n \left(\frac{r}{2}\right)^i
[\gamma + \sqrt{-mE}]^{i+1},
\label{eq:Sn}
\eeq although actually at $n=3$ and beyond contributions from the
shape parameter term, and other terms omitted in Eq.~(\ref{eq:TBamp}),
need to be added in order to get the full result to $O(R^n/a_2^n)$.
Regardless, once $n > 0$, the $\tau$ of Eq.~(\ref{eq:Sn}) no longer
has the same residue at $E=-B_2$. The relationship between the
K-matrix and the phase shifts is affected by this change in
wave-function renormalization. It becomes: \beq
K^{(n)}(k,k;E)=-\frac{1}{\sum_{i=0}^n (\gamma r)^n} \frac{3m}{8\gamma
  k}\tan\delta.
\label{eq:Kn}
\eeq

Since $K^{(n)}(0,0;-B_2)$ enters the driving term of our subtracted
integral equation for the threshold amplitude, not
only the kernel, but also the inhomogeneous term, of the integral
equation, changes as we go to higher order. This, however, is the only
change in the driving term of that equation, since ${\cal Z}$ is the
same no matter what order in $R/a_2$ we calculate to.

Now that we have inserted the additional two-body length scale $r \sim
R$ into our theory, the breakdown scale of the EFT is explicit in our three-body
integral equations. In order to demonstrate renormalization-group
invariance of observables in the theory we must show that amplitudes
for $p \ll 1/r$ are cutoff independent (up to higher-order terms) for
$\Lambda \gg 1/r$. Therefore in what follows we analyze the integral
equation for the $n$th-order amplitude in 
the domain $p \ll 1/r \ll \Lambda$. 

\begin{figure}[tb]
\centerline{\includegraphics*[width=12cm,angle=0]{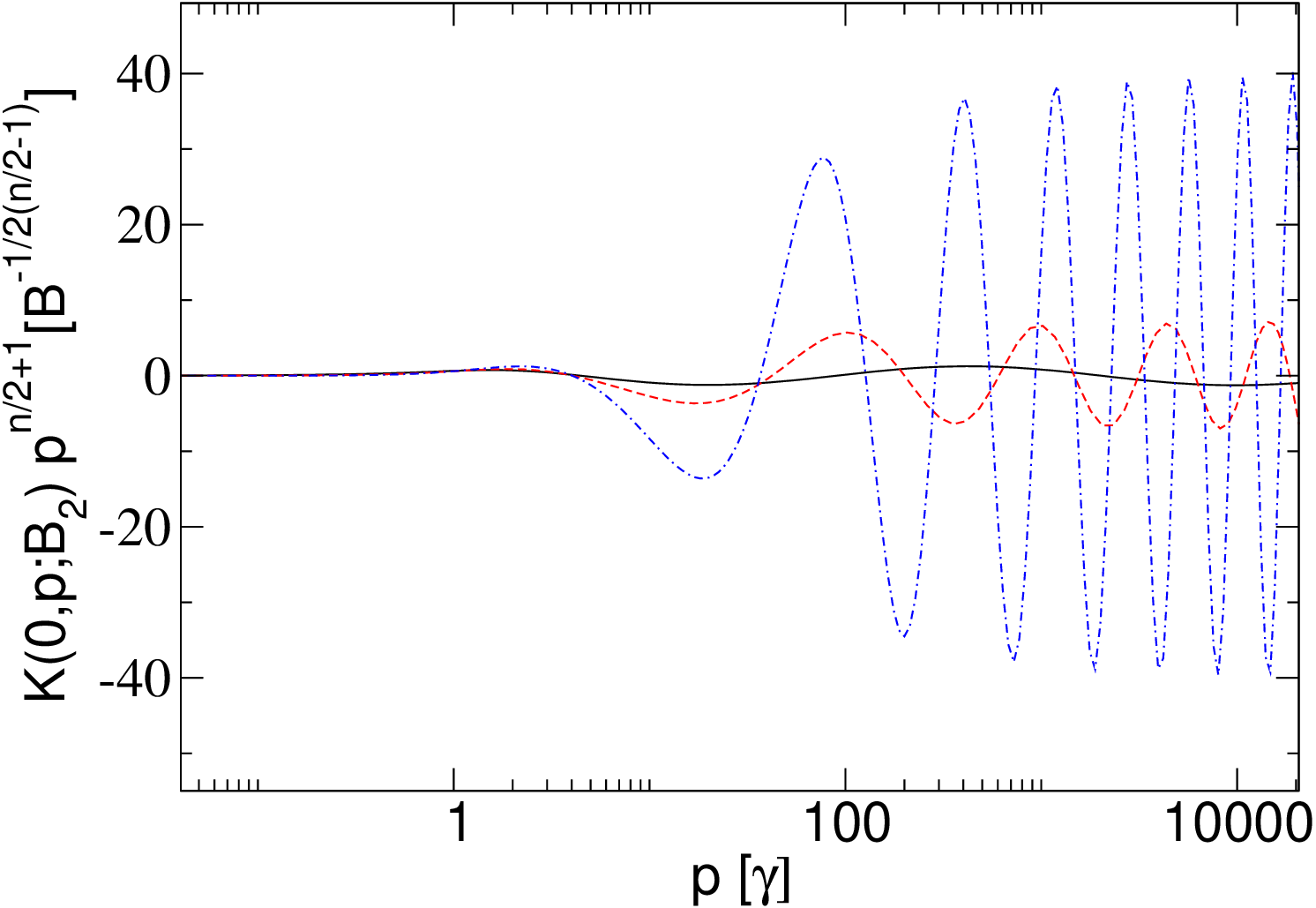}}
\caption{\label{fig:k0p} The amplitude $K^{(n)}(p,0;-B_2)p^{n/2+1}$ as
a function of the momentum $p$ with $r=0.073 \gamma^{-1}$,
$a_3=1.362 \gamma^{-1}$ and $\Lambda=10000 \gamma$. The solid, dashed, and dot-dashed lines
denote the LO, NLO, NNLO result, respectively. The LO line reproduces
the results given in \cite{Bedaque:1998kg,Afnan:2003bs}.}
\end{figure}

Consider the integral equation at threshold at NLO ($n=1$) or NNLO ($n=2$):
\bea
K^{(n)}(p,0;-B_2)&=&\frac{2 m}{p^2+\gamma^2}\nonumber\\
&& \quad +\, \frac{4
  m^2}{3}\int_0^\Lambda \hbox{d}q\frac{1}{ p q} \log\left(\frac{p^2+q^2+p
  q+\gamma^2}{p^2+q^2-p q+\gamma^2}\right) S^{(n)}(-B_2;q)K^{(n)}(q,0;-B_2), 
\nonumber\\
\eea
with
\begin{equation}
S^{(n)}(-B_2;q) \equiv S^{(n)} \left(-B_2 - \frac{3q^2}{4m}\right),
\end{equation}
where the right-hand side of this equation is given by
(\ref{eq:Sn}). For $p \gg \gamma$, and
after a subtraction, this equation becomes
\bea
\label{eq:rganalysisstart}
K^{(n)}(p,0;-B_2)&=&-\frac{2 m}{\gamma^2}+K^{(n)}(0,0;-B_2)
\nonumber\\[0.3cm]
&&\quad+\, \frac{4m^2}{3}\int_0^\Lambda\hbox{d}q\biggl[\frac{1}{p q}
\log\left(\frac{p^2+p q+q^2}{p^2-p q+q^2}\right)-\frac{2}{q^2}\biggr]
S^{(n)}(-B_2;q)K^{(n)}(q,0;-B_2).\nonumber\\
\label{eq:IEhighp}
\eea
We must now consider two different momentum domains within the
integral on the right-hand side of Eq.~(\ref{eq:IEhighp}).
In the first we have
momenta  $q\ll 1/r$, which are within the domain of validity of
our EFT,
while in the second we have
$q\gg 1/r$, i.e. $q$ is above the EFT breakdown scale.
To make this transparent we separate the integral in
Eq.~(\ref{eq:rganalysisstart}):
\bea
K^{(n)}(p,0;-B_2)&=&\frac{m \alpha}{\gamma^2}+\frac{4m^2}{3}\int_0^{1/r}
\biggl[\frac{1}{pq}\log\left(\frac{p^2+q p+q^2}{p^2-q p+q^2}\right)
-\frac{2}{q^2}\biggr]
S^{(n)}(-B_2;q)K^{(n)}(q,0;-B_2)\nonumber\\[0.3cm]
&&\quad +\frac{4 m^2}{3}\int_{1/r}^\Lambda\hbox{d}q
\biggl[\frac{1}{pq}\log\left(\frac{p^2+q p+q^2}{p^2-q p+q^2}\right)
-\frac{2}{q^2}\biggr]
S^{(n)}(-B_2;q)K^{(n)}(q,0;-B_2),\nonumber\\
\eea
where $\alpha$ is a number of order one:
\beq
\alpha=\frac{3 a_3  \gamma}{8} - 2.
\eeq
Now we take $p \ll 1/r$, giving:
\bea
K^{(n)}(p,0;-B_2)&=&\frac{m \alpha}{\gamma^2}+\frac{4 m^2}{3}\int_0^{1/r}
\hbox{d}q\biggl[\frac{1}{p q}
\log\left(\frac{p^2+q p+q^2}{p^2-q p+q^2}\right)-
\frac{2}{q^2}\biggr]S^{(n)}(-B_2;q)K^{(n)}(q,0;-B_2)\nonumber\\[0.3cm]
&&\qquad - \frac{16
  m^2}{9}\int_{1/r}^\Lambda\hbox{d}q\,\frac{p^2}{q^4} \, 
S^{(n)}(-B_2;q)K^{(n)}(q,0;-B_2)~.
\eea
For $q \gg 1/ r$ we have
\beq
S^{(n)}(-B_2;q) 
\approx\frac{3^{1/2} q}{\pi m^2} \left(\frac {\sqrt{3} r q}{4}\right)^n \; ,
\eeq
and we can reduce the second integration to obtain
\bea
K^{(n)}(p,0;-B_2)&=&\frac{m \alpha}{\gamma^2}
+\frac{4 m^2}{3}\int_0^{1/r}\hbox{d}q
\biggl[\frac{1}{p q}\log\left(\frac{p^2+p q+q^2}{p^2-p q+q^2}\right)
-\frac{2}{q^2}\biggr]S^{(n)}(-B_2;q)K^{(n)}(q,0;-B_2)\nonumber\\[0.3cm]
&&\qquad\qquad\qquad\qquad- p^2 \left(\frac{\sqrt{3} r}{4}\right)^n 
\frac{16}{3 \sqrt{3} \pi} \int_{1/r}^\Lambda
\hbox{d}q \frac{K^{(n)}(q,0;-B_2)}{q^{3-n}}~.
\eea
For $q\gg1/r$ we observe numerically that:
\begin{equation}
K^{(n)}(q,0;-B_2)=\frac{m \beta^{(n)}(q)}{q^{1 + n/2}},
\label{eq:Knfalloff}
\end{equation}
with $\beta^{(n)}(q)$ oscillatory and bounded. (See Fig.~\ref{fig:k0p}.)
Therefore for $p\ll1/r$ we can write
the integral equation as
\bea
K^{(n)}(p,0;-B_2)&=&\frac{m \alpha}{\gamma^2}+\frac{4 m^2}{3}\int_0^{1/r}
\hbox{d}q
\biggl[\frac{1}{p q}\log\left(\frac{p^2+p q+q^2}{p^2-p q+q^2}\right)-
\frac{2}{q^2}\biggr]
S^{(n)}(-B_2;q)K^{(n)}(q,0;-B_2)
\nonumber\\
&&\qquad\qquad-m p^2 
\left(\frac{\sqrt{3} r}{4}\right)^n 
\frac{16}{3 \sqrt{3} \pi} \int_{1/r}^\Lambda
\hbox{d}q \, \frac{\beta^{(n)}(q)}{q^{4 - n/2}}~.
\eea

The second integral converges even when the cutoff is taken to
infinity. Indeed, for $\Lambda \gg 1/r$ it is dominated by $q \sim
1/r$, which allows us to estimate its size to be:
\begin{eqnarray*}
\int_{1/r}^\Lambda
\hbox{d}q \, \frac{\beta^{(n)}(q)}{q^{4-n/2}}
\sim r^{3 - n/2} \, \beta^{(n)}\left(\frac{1}{r} \right).
\end{eqnarray*}
Therefore, up to corrections
of this size, the high-momentum piece of the integral equation
decouples and
\begin{equation}
K^{(n)}(p,0;-B_2) \sim \frac{m}{\gamma p}~\mbox{for}~\gamma \ll p \ll
1/r,
\label{eq:pgggamma}
\end{equation}
for $n=1$ and $n=2$, which is
the same behavior for $p \gg \gamma$ as at leading order.
Matching the low-momentum behavior of Eq.~(\ref{eq:pgggamma}) to the
observed high-momentum behavior (\ref{eq:Knfalloff}) allows us to
determine that:
\begin{equation}
\beta^{(n)}\left(\frac{1}{r}\right) \sim
\left(\frac{1}{r}\right)^{n/2} \frac{1}{\gamma}.
\end{equation}
This finally leads us to conclude that the size of the high-momentum
piece ($q \gg 1/r$) of the homogeneous term in the integral equation
is $\sim \frac{m}{\gamma} p^2 r^3$, as long as $p \ll 1/r$. Comparing
to the behavior of Eq.~(\ref{eq:pgggamma}) we see that the
contribution of high-momentum modes is suppressed by three powers of
the small parameter $p r$, and so will only be significant if one
wishes to do a calculation up to N$^3$LO accuracy.  At both
next-to-leading and next-to-next-to-leading order the threshold
amplitude obtained by the subtraction technique will be cutoff
independent up to terms which are higher order in $r/a_2$ and/or $pr$.

In fact, the behavior (\ref{eq:Knfalloff}) of $K^{(n)}(q,0;-B_2)$ at large
$q$ can be understood via Hilbert-Schmidt theory. If we first
symmetrize the kernel of our (unsubtracted) integral equation, and
then make a Hilbert-Schmidt decomposition of the new
kernel~\footnote{Since the procedure for the subtraction is carefully
  designed to maintain the symmetry of the K-matrix the subtraction
  does not modify the validity of the Hilbert-Schmidt decomposition.},
we find that the solution for the threshold amplitude can be written
as:
\begin{equation}
K^{(n)}(q,q';-B_2)=\frac{\Gamma_m^{(n)}(q)
  \Gamma_m^{(n)}(q')}{\lambda_m}.
\label{eq:HS}
\end{equation}
In Eq.~(\ref{eq:HS}) $\{\Gamma_m:m=1,\ldots,\infty\}$ is
the set of eigenfunctions of the
integral-equation kernel, with each one corresponding to a different
eigenvalue $\lambda_m$.
These eigenfunctions are only normalizable 
in the $\Lambda \rightarrow \infty$ limit if
\begin{equation}
\int_0^\infty \hbox{d}q \, \Gamma_m^{(n)}(q) S^{(n)}(q) \Gamma_m^{(n)}(q) < \infty.
\label{eq:normalize}
\end{equation}
In the case of the homogeneous equation which predicts the
eigenenergies at which $1+2$ bound states appear the requirement
(\ref{eq:normalize}) corresponds to demanding that the bound-state
wave function is normalizable.

Assuming that the eigenfunction that is largest at high momentum falls
off as $\Gamma^{(n)} \sim \frac{1}{q^{\alpha_n}}$ and using the
general asymptotic behavior of $S$ at arbitrary order $n$: $S^{(n)}
\sim q^{n+1}$, Eq.~(\ref{eq:normalize}) implies:
\begin{equation}
n + 1 - 2\Re(\alpha_n) \leq -1 \Rightarrow \Re(\alpha_n) \geq
\frac{n}{2} + 1.
\label{eq:falloff}
\end{equation}
Substituting this behavior back into the Hilbert-Schmidt decomposition
of $K^{(n)}$, Eq.~(\ref{eq:HS}), we see that $K^{(n)}(q,q';-B_2)$ must
fall off at least as fast as $\frac{1}{q^{n/2 + 1}}$ for $q', \gamma \ll
q$. Numerically we find that the equality in Eq.~(\ref{eq:falloff}) is
realized at LO, NLO, and NNLO. But as we have seen previously, such a
fall-off with off-shell momentum $q$ is already enough to guarantee
the decoupling of the high-momentum ($q \gg 1/r$) modes in the integral
equation that governs the threshold scattering amplitude.

We now turn our attention to the amplitude when $E > -B_2$. 
The NLO and NNLO versions of
Eqs.~(\ref{eq:full-amplitude-subtracted})--(\ref{eq:deltatau}) are
  found by making by the replacements:
\begin{eqnarray}
K^{(0)}(q,q'';-B_2) &\longrightarrow& K^{(n)}(q,q'';-B_2)\\
\tau^{(0)}(E;q) &\longrightarrow& \tau^{(n)}(E;q),
\end{eqnarray}
The question is whether these equations yield cutoff-independent
predictions for the amplitude $K(p,q;E)$ when $p,q,k \ll 1/r$.

Since the fall-off of $\delta {\cal Z}$ at large momenta 
is, {\it at worst}, that given in Eq.~(\ref{eq:deltaZ}) we see that
as $q \rightarrow \infty$, $\delta {\cal Z}(q',q;E)$
falls off at least as fast as $\Gamma_m^{(n)}(q)$---at least for $n=1$
and $n=2$, which are the only cases we consider here. In consequence, as
long as the integral:
\begin{equation}
\int_0^\Lambda \hbox{d}q' \, q'^2 \, \Gamma_i^{(n)}(q') \tau^{(n)}(E;q')
\Gamma_j^{(n)}(q')
\label{eq:key}
\end{equation}
is not cutoff-dependent then
Eqs.~(\ref{eq:full-amplitude-subtracted})--(\ref{eq:deltatau}) will
yield cutoff-independent results. But we already know from
Eq.~(\ref{eq:normalize}) that this integral is convergent in the
$\Lambda \rightarrow \infty$ limit, and so we
are guaranteed that, at least at NLO and NNLO, the formalism of the
previous section can be adapted so that, at any energy such that 
$k \ll 1/r$, it
yields cutoff-independent results
for the low-momentum amplitude.

We note, in passing, that the integral (\ref{eq:key}) would {\it not}
be convergent if $\Gamma_m^{(n)}(q) \sim 1/q$ as $q \rightarrow
\infty$, i.e. the NLO and NNLO amplitudes had the same fall-off at $q
\gg 1/r$ as the
LO amplitude.  If the LO high-momentum behavior were also present in
the higher-order cases then the integral of Eq.~(\ref{eq:key}) would
behave as $r^2 k^2 \ln \Lambda$ for large $\Lambda$. The low-momentum
amplitude would then not be renormalization-group invariant, and
a counterterm, like the one introduced in Ref.~\cite{Bedaque:2002yg},
would be necessary to remove this divergence. However, our
numerical results (see Section~\ref{sec:results})
indicate {\it no} $\Lambda$-dependence (to six
significant figures) in the NNLO phase shifts, because the fall-off of
$K^{(n)}(q,0;-B_2)$ with $q$ is really $1/q^{1+n/2}$ and not $1/q$.

Therefore we have shown that with the changes in $S$ written in
Eq.~(\ref{eq:Sn}), and the corresponding changes in the inhomogeneous
term, we can use Eqs.~(\ref{eq:threshsubt}) and
(\ref{eq:full-amplitude-subtracted})--(\ref{eq:deltatau}) to compute
three-body scattering and bound-state observables at NLO and NNLO.  In
these equations we can take the cutoff to be arbitrarily large. Therefore we have
proven---using a combination of numerical and analytic
arguments---that 
no new counterterm is needed to
renormalize the three-body system with short-range interactions at
NNLO.  This conclusion is driven by the behavior of the zero-energy
scattering amplitude for off-shell momenta $q \gg 1/r$, and that
behavior is tied to the requirement that the bound-state wave
functions of the $1+2$ system be normalizable.
\section{The Three-Boson System at NNLO}
\label{sec:results}

As mentioned above, few-body systems of $^4$He atoms seem to be an 
ideal testing ground for the formalism laid out here. Helium-4
clusters are bound by the van der Waals force, a potential with a
long-range tail of the form $-C_6/r^6$.  The S-wave phase shifts for
this potential obey a modified effective-range expansion: $k \cot
\delta$ can be expanded in powers of $k^2$, but non-polynomial terms
appear at order $k^4$. As pointed out by Braaten and Hammer~\cite{Braaten:2002jv,Braaten:2004rn}, this means that an EFT with
contact interactions should be able to mimic the underlying
van der Waals interaction up to this order, even though the non-analytic
effects at $O(k^4)$ cannot be reproduced by a short-range potential.
The size of the $O(k^2)$ term in the modified effective-range
expansion is $l \approx (m C_6/\hbar^2)^{1/4}$, and so, when applied
to these systems, the EFT with contact interactions is an expansion in
$l/a_2$ and $kl$ instead of $R/a_2$ and $kR$.  In the following we will
present numerical results for the $^4$He three-body system up to 
NNLO in this expansion.

The atom-atom scattering length
$a_2=\left(104^{+8}_{-18}\right)$~\AA\ and the dimer binding energy
$B_2=1.30962$ mK were derived from measurements of the dimer bond
length, and using the zero-range approximation \cite{Gri00} this
gives an effective range of the order of 10 \AA.
Therefore the EFT expansion parameter is $l/a_2 \approx 0.1$ and
the EFT expansion should converge rather quickly. Since we compute observables
up to NNLO we expect our results at that order have a
remaining error $\sim 0.1$\%.

To compute observables in the three-body system we need one three-body
datum. However, the experimental information on clusters of $^4$He
bound states with more than two atoms is rather limited.  While the
$^4$He trimer and larger clusters have been observed, no quantitative
information about their binding energies is available
\cite{STo96,BST02}. On the other hand, various ``realistic''
potentials have been developed (see, e.g.~\cite{TTYref,HFDBref}) and
have been employed for few-body calculations
\cite{NFJ98,RoY00,Ro00,MSSK01,Barl02}.  In need of a three-body
observable to fix the value of our three-body parameter we will employ
these calculations as our three-body input.

\begin{table}[t]
\begin{center}
\begin{tabular}{|c|c|c|c|c|}
\hline
Input      &      &$B_3^{(1)}[B_2$]  & $B_3^{(0)}[B_2]$  &
$a_3[\gamma^{-1}]$\\
\hline\hline%
$B^{(1)}_3$& LO  & 1.738  & 99.27  &  1.179\\
           & NLO & 1.738  & 84.87  &  1.199\\
           & NNLO& 1.738  & 89.52  &  1.203\\
\hline
$a_3$   & LO   & 1.723  & 97.12  &  1.205\\
           & NLO  & 1.735  & 84.48  &  1.205\\
           & NNLO & 1.737  & 89.38  &  1.205\\
\hline\hline
TTY\cite{RoY00,Ro00}&       & 1.738  & 96.33  &  1.205\\
\hline
\end{tabular}
\end{center}
\caption{\label{table:ttyroudnev}
EFT predictions for the $^4$He trimer binding energies and
atom-dimer scattering length $a_3$ up to N$^2$LO. 
Energies and lengths are given in units of two-body binding
energy. The first three rows show LO, NLO, and NNLO results
when $B_3^{(1)}$ is used as input. The next three rows give the
results when $a_3$ is chosen as input.
In the last row we display the results of the calculations of
Ref.~\cite{RoY00,Ro00}.}
\end{table}
\begin{table}[t]
\begin{center}
\begin{tabular}{|c|c|c|c|c|}
\hline
Input      &      &$B_3^{(1)}[B_2$]  & $B_3^{(0)}[B_2]$  &
$a_3[\gamma^{-1}]$\\
\hline\hline%
$B^{(1)}_3$& LO   & 1.741  & 99.78  &  1.173\\
           & NLO  & 1.741  & 85.31  &  1.193\\                  
           & NNLO & 1.741  & 90.08  &  1.196\\
 \hline

$a_3$   & LO   & 1.643  & 85.02  &  1.362\\
           & NLO  & 1.655  & 74.79  &  1.362\\
           & NNLO & 1.656  & 78.82  &  1.362\\
               
\hline\hline
TTY\cite{MSSK01}  &        & 1.741  & 96.06  &  1.362\\
\hline
\end{tabular}
\end{center}
\caption{\label{table:ttymoto}
EFT predictions for the $^4$He trimer binding energies and
atom-dimer scattering length $a_3$ up to N$^2$LO. 
Energies and lengths are given in units of two-body binding
energy. The first three rows show LO, NLO, and NNLO results
when $B_3^{(1)}$ is used as input. The next three rows give the
results when $a_3$ is chosen as input.
In the last row we display the results of the calculations of
Ref.~\cite{MSSK01}.}
\end{table}

This strategy follows that of 
Braaten and Hammer \cite{Braaten:2002jv}, who
analyzed universal properties
of three-body systems of $^4$He atoms using the EFT with contact
interactions alone. They computed various observables,
including the trimer energies and the atom-dimer scattering lengths, using
the results obtained with different atom-atom potentials
as input parameters. They achieved the desired accuracy of 10\% for
a leading-order EFT calculation and estimated the NLO corrections for low-energy
observables by assuming they would shift the leading-order
result by a factor proportional to $l/a_2$.

The ``TTY potential''~\cite{TTYref} is one of the potential models
used to predict the binding energies of clusters of ${}^4$He atoms.
As with all such models, it relies on theoretical assumptions about
the underlying atom-atom interactions.  It reproduces the experimental
value of the dimer binding energy given above. The effective range has
been calculated for various potentials by Janzen and Aziz, and for the
TTY potential they obtain $r=7.329$~\AA~\cite{jan00}.  When used in a
three-body calculation the TTY potential produces exactly two
three-body bound states. There have been several calculations of these
trimer binding energies, $B_3^{(0)}$ and $B_3^{(1)}$, employing the
TTY potential, and all of them agree on those energies to within
0.5\%.

The situation for the atom-dimer scattering length is somewhat less
clear. Two recent Faddeev calculations which both employ the TTY
potential agree on the trimer binding energies, but have values for
$a_3$ which disagree at the 10\% level~\cite{MSSK01,Ro00}.  This is
more than the combined errors stated in Ref.~\cite{MSSK01} (5~\AA
$\approx$ 4\%) and Ref.~\cite{Ro00} (0.1~\AA $\approx$ 0.1\%). In what
follows we will examine both Faddeev calculations in light of the EFT
treatment developed in the previous sections. Although we focus
largely on the results of these calculations when the TTY potential is
their two-body input, the pattern of results that we find is the same
if we compare to calculations that employ other atom-atom potentials
instead.

Our results for the TTY potential using input from the computation of
Refs.~\cite{RoY00,Ro00} are shown in Table~\ref{table:ttyroudnev}.
Note that for better comparison we have computed all bound state
observables in units of the two-body binding energy $B_2$.  At low
orders the results change somewhat depending on whether we use the
atom-dimer scattering length or the energy of the higher-lying trimer
$B_3^{(1)}$ as the three-body input datum, and so in the table we
present the numbers we obtained via both strategies.  In this
case the dependence on the choice of input becomes systematically
less as the calculation is carried to higher orders---one sign that
the EFT is converging in a self-consistent way.  Our computations are
accurate to at least the number of digits shown in the table.  The
results are also cutoff independent for sufficiently large cutoffs,
which supports the claim of the previous section that no additional
three-body counterterm is needed for the renormalization of the
three-boson problem at NNLO. 

At NNLO we achieve very good agreement for the energy of the excited
three-body bound state or the atom-dimer scattering length if we use
(respectively) $a_3$ or $B_3^{(1)}$ as calculated in \cite{Ro00}.
If we use $B_3^{(1)}$ as the input parameter we obtain $a_3=1.203
\gamma^{-1}$, to be compared to $1.205 \gamma^{-1}$ obtained by
Roudnev. The remaining discrepancy---which is significant given the
accuracy of our calculation and of that of Ref.~\cite{Ro00}---is about
0.1\%, exactly what we expected from a calculation at this order. The
situation for the deep-lying trimer is less pleasing. We obtain $89.38
B_2$ for the three-body ground state, which has to be compared to 
Roudnev's
full Faddeev TTY-potential result of $96.33 B_2$.  The disagreement
between these results for the lowest-lying three-body bound state
appears to be due to slow convergence of the EFT for that
observable. This binding energy changes by roughly 15\% from LO to NLO
and 5\% from NLO to NNLO.  (The fact that the NLO correction is large,
and moves the EFT result {\it away} from the calculation of
Ref.~\cite{MSSK01} has also been found in an EFT calculation that uses
the strictly perturbative $l/a_2$ expansion~\cite{Hammerunp}.)  Since 
$B_3^{(0)}$ is large on the scale of $B_2$
this
relatively slow convergence is not surprising, 
although the pattern of NLO and NNLO
corrections suggests that an N$^3$LO calculation of $B_3^{(0)}$ should 
have at worst a 2\% error.
 
Since Roudnev \& Yakovlev and Motovilov {\it et al.} essentially agree
on the value of $B_3^{(1)}$, but disagree on the number for $a_3$, the
fact that we reproduce the $a_3$ result of Ref.~\cite{Ro00} means that
we cannot also reproduce the numbers of Motovilov {\it et al.}.  This
can be seen explicitly in Table \ref{table:ttymoto}. If we use, for
example, their result for $B_3^{(1)}$ as our three-body input we
obtain $1.196 \gamma^{-1}$ for the atom-dimer scattering length, as
compared to their $1.362 \gamma^{-1}$.  In contrast to the result for
$B_3^{(0)}$ the EFT result for $a_3$ shown in
Table~\ref{table:ttymoto} is converged at NNLO, but it does not
reproduce the number quoted in Ref.~\cite{MSSK01}.  We see this as a
strong indication that the atom-dimer scattering lengths obtained by
Motovilov {\it et al.}  have a larger error than the 5 \AA\ quoted by
the authors.

\begin{figure}[t]
\centerline{\includegraphics*[width=12cm,angle=0]{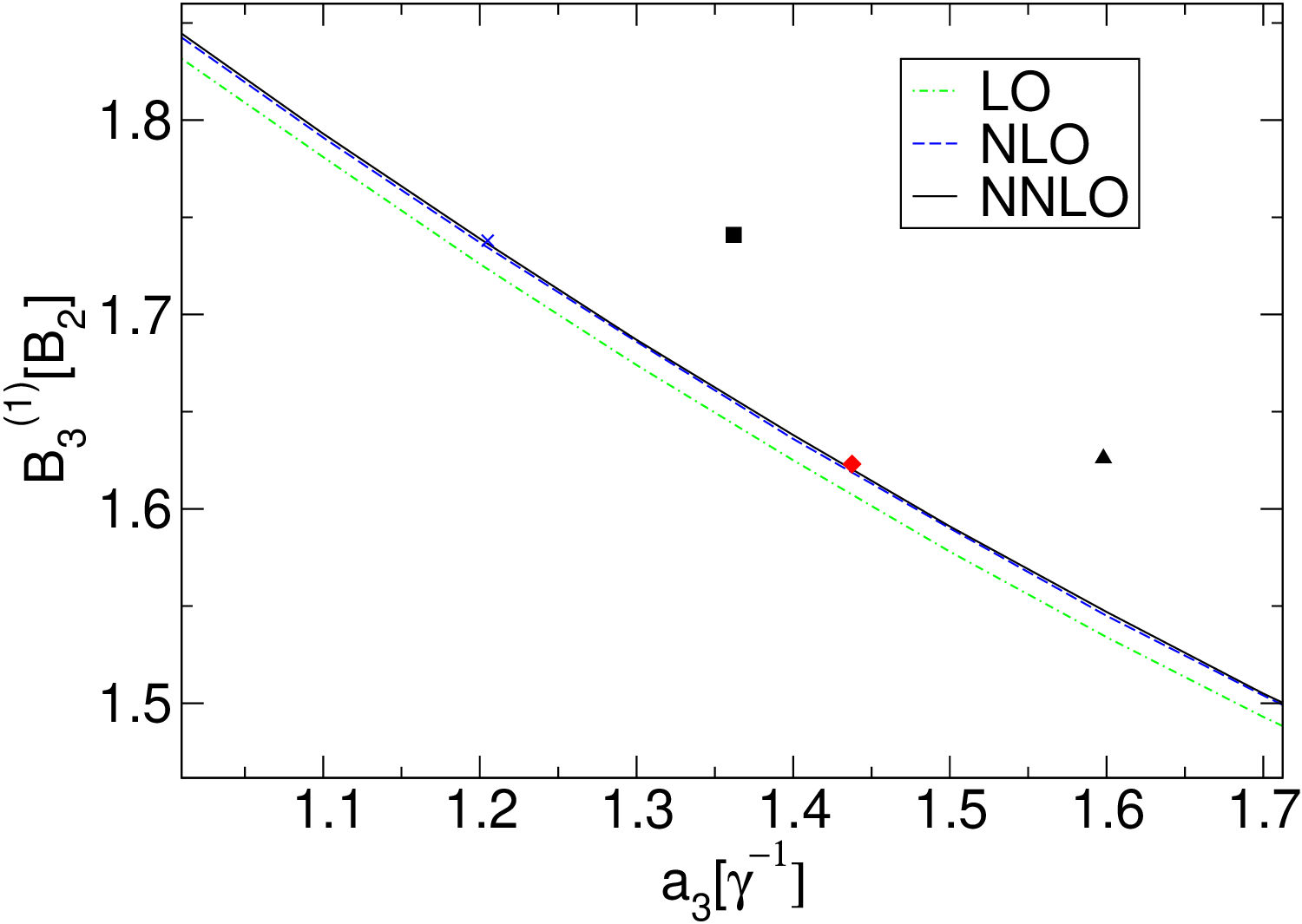}}
\caption{\label{fig:phillipsline} The Phillips line for leading (dot-dashed line),
next-to-leading (dashed line)
and next-to-next-to-leading (solid line) order. The cross and diamond correspond to the
TTY and HFD-B prediction obtained in \cite{RoY00,Ro00}, respectively.
The square and triangle denote the corresponding predictions
obtained in \cite{MSSK01}.}
\end{figure}

A further way to illustrate this point is to consider the Phillips
line. The Phillips line is a universal feature in three-body systems
with large two-body scattering lengths: it is a nearly linear
correlation between the 1+2 scattering length and the
three-body binding energy.  In Fig.~\ref{fig:phillipsline} we display
the change of the Phillips line with the inclusion of higher-order
corrections. The Phillips line is converging to a definite result for
a particular value of $r/a_2$. In Fig.~\ref{fig:phillipsline} we also
display the data points for the TTY potential and the HFD-B potential
as given in \cite{MSSK01} and \cite{RoY00,Ro00}
\footnote{The effective range of the HFD-B potential is 7.28
  \AA\ \cite{jan00} and therefore differs by 1\%  from the TTY
  effective range. However, this difference amounts to a 0.1\% shift
  in observables which is negligible on the scale of this plot.  }. It
is clear that the values obtained by Roudnev and Yakovlev lie very
close to the line, while the values obtained by Motovilov {\it et al.}
seem to be systematically shifted away from it. This appears to
support Roudnev's argument in Ref.~\cite{Ro00} that the $a_3$ values of
Motovilov {\it et al.} were not calculated on a large enough spatial
grid. Very accurate phase-shift calculations at energies close to the
atom-dimer threshold are required for a precise determination of $a_3$.

\begin{figure}[tb]
\centerline{\includegraphics*[width=12cm,angle=0]{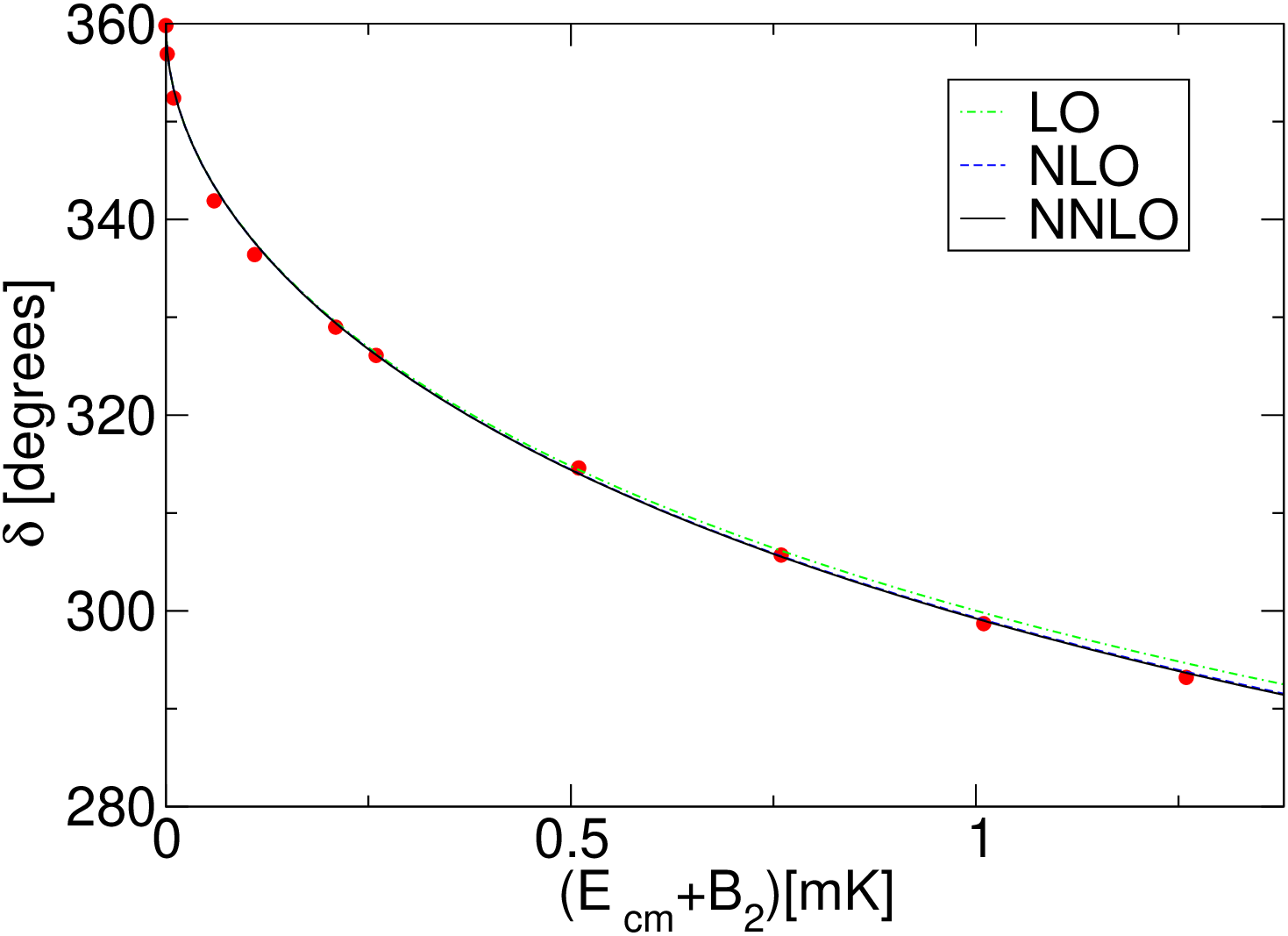}}
\caption{\label{fig:kcotd}The LO (dot-dashed line), NLO (dashed line), and
NNLO (solid line) S-wave scattering phase shifts $\delta_0$
using the TTY results for $B_2$ and $r$ and the TTY
prediction of Ref.~\cite{MSSK01} for the atom-dimer scattering length. The filled circles show the results for $\delta_0(E)$ given in  \cite{MSSK01}.
}
\end{figure}

We also computed the phaseshift $\delta_0$ with $a_3=1.36
\gamma^{-1}$ as the three-body input for our subtraction. In
Fig.~\ref{fig:kcotd} we display our results for LO, NLO, and NNLO in
each case and compare it to the TTY results given in
Refs.~\cite{MSSK01}. In this figure we display Motovilov {\it et al.}'s phase shifts
because the criticism leveled by Roudnev at their
calculation should become less of an issue as the energy increases
away from the two-body threshold.

There are a few interesting points to note regarding
Fig.~\ref{fig:kcotd}. First, by NNLO the EFT seems to have converged
to a definite result for $\delta_0(E)$ in this range---once $a_3$ is
fixed.  Second, the NNLO EFT results agree well with
the results of the Faddeev calculations using the phenomenological TTY
potential.  The use of the atom-dimer scattering length
obtained in \cite{MSSK01} as the three-body input accounts for
whatever difficulties they may have had in computing phase shifts
close to two-body threshold.
 Third, the NLO and NNLO corrections definitely make the
energy-dependence of the EFT results significantly closer to the
energy-dependence of the TTY-potential phase shifts. Since the slope
of the phase shift at zero energy is fixed to be $a_3$, and so is
independent of the order of the calculation, it is not surprising that
the effect of the higher-order corrections is most noticeable at the
higher energies shown.

\section{Summary}
\label{sec:resdis}
In this work we have extended the subtraction formalism developed
previously \cite{Hammer:2000nf,Afnan:2003bs} to higher orders in the
$R/a_2$ expansion.  Using analytical and numerical arguments we have
shown that the three-body system at NNLO can be renormalized without
the need for a second three-body datum.  Therefore, one does not need
to include an energy-dependent three-body force at this order if one
accepts renormalizability and naive dimensional analysis as the
guidelines for the establishment of a consistent power counting. This
is an interesting result as it provides another example of the
potential predictive power of the EFT with contact interactions alone.

Few-body systems of spinless bosons like $^4$He atoms are the
simplest testing ground for this EFT.
We have advanced the leading-order EFT calculation by Braaten and
Hammer by two orders---to NNLO---and found interesting results.

Without a three-body input at leading order the EFT with contact
interactions alone is not able to deliver any information about the
three-body system.  Thus, we used a three-body observable calculated
with a specific phenomenological inter-atomic potential (``the TTY
potential'') to fix our three-body input.  Using two two-body inputs
and one three-body input we achieved very good agreement with the
Faddeev calculations of Roudnev and Yakovlev \cite{RoY00,Ro00} for the TTY potential.
The disagreement between our calculation and the results
obtained by Motovilov {\it el. al} is of the order of 10\%, which
is two orders of magnitude larger than the expected accuracy of an
NNLO calculation. 
Our calculation therefore supports the claim of Ref.~\cite{Ro00} that
there were problems with the numbers quoted in Ref.~\cite{MSSK01}, and
that the error given there should be enlarged. We
believe that this is an example of EFT's ability
to check the consistency of different
phenomenological calculations. The numbers of Ref.~\cite{MSSK01} for
$a_3$ are simply not consistent with the model-independent pattern
of $r/a_2$ corrections to the Phillips line that we have calculated
using the short-range EFT.

Furthermore, it is worth noting that although various inter-atomic
potentials give similar results for low-energy observables in the
two-body system, they give different results for three-body
observables---albeit in correlated ways that are constrained by
relations like that shown in Fig.~\ref{fig:phillipsline}. But, to
decide which two-body potential gives the most appropriate description
of the three-body system, an experimental measurement of at least one
three-body observable is needed.  The measurement of one such number
(e.g. $B_3^{(1)}$) would therefore be a valuable test of the current
realistic potentials and would further our understanding of the
atom-atom interaction.  The results of this paper show that once one
knows, for instance, $B_3^{(1)}$, the EFT can be used to generate
precise, model-independent, predictions for remaining low-energy
three-body observables.

A further application of the formalism presented in this work is the
three-nucleon sector. The nucleon-nucleon force is of finite range and
so there the effective-range expansion is analytic to all orders in
$k^2$.  We have seen here that the short-range EFT converges well for
systems of three Helium-4 atoms, and it has been shown that this EFT
also converges as expected in the two-nucleon sector, where some
observables have been computed up to four orders in the $R/a_2$
expansion~\cite{Rupak,CBK}. The extension of the equations developed
here to the the three-nucleon problem requires only the addition of
spin and isospin degrees of
freedom~\cite{Afnan:2003bs,Bedaque:1999ve}. This will open the way to
EFT computations of low-energy three-body observables (including those
involving electroweak probes) which have errors of $< 5$\%.

\begin{acknowledgments}
We thank Eric Braaten and Hans-Werner Hammer for useful
discussions. We also thank Iraj Afnan for his comments on the
manuscript. One of us (D.~R.~P.) acknowledges the hospitality of Chubu
University's Department of Natural Science, where this work was completed.
This work was supported by the U.S. Department of Energy
under grant DE-FG02-93ER40756.

\end{acknowledgments}

\end{document}